\documentclass[dvipsnames, dvipsnames, sigconf]{acmart}
\AtBeginDocument{%
  \providecommand\BibTeX{{%
    \normalfont B\kern-0.5em{\scshape i\kern-0.25em b}\kern-0.8em\TeX}}}
\usepackage{xcolor}
\usepackage{graphicx,tabularx,subcaption}
\usepackage{booktabs}
\usepackage{multirow}
\usepackage[capitalise,nameinlink,compress]{cleveref}
\usepackage{comment}
\usepackage[english]{babel}
\usepackage{fontawesome5}
\usepackage{arydshln}
\usepackage{pdfpages}

\graphicspath{{images/}}

\definecolor{study1}{RGB}{235, 255, 245}
\definecolor{study1font}{RGB}{0, 178, 120}

\definecolor{study2}{RGB}{231, 239, 253}
\definecolor{study2font}{RGB}{19, 99, 223}

\definecolor{study3}{RGB}{241, 232, 248}
\definecolor{study3font}{RGB}{112, 48, 160}

\definecolor{futurework}{RGB}{252, 230, 208}
\definecolor{futureworkfont}{RGB}{249, 111, 7}

\copyrightyear{2024} 
\acmYear{2024} 
\setcopyright{rightsretained} 
\acmConference[CHI EA '24]{Extended Abstracts of the CHI Conference on Human Factors in Computing Systems}{May 11--16, 2024}{Honolulu, HI, USA}
\acmBooktitle{Extended Abstracts of the CHI Conference on Human Factors in Computing Systems (CHI EA '24), May 11--16, 2024, Honolulu, HI, USA}
\acmDOI{10.1145/3613905.3638177}
\acmISBN{979-8-4007-0331-7/24/05}

\definecolor{main1}{HTML}{00B278}    
\definecolor{sub1}{HTML}{EBFFF5}     

\definecolor{main2}{HTML}{1363DF}    
\definecolor{sub2}{HTML}{E7EFFD}     

\definecolor{main3}{HTML}{7030A0}    
\definecolor{sub3}{HTML}{F1E8F8}     

\usepackage{framed}
\colorlet{shadecolor}{futurework}
\colorlet{framecolor}{futureworkfont}

\newenvironment{frshaded*}{%
\MakeFramed {\advance\hsize-\width \FrameRestore}
}%
{\endMakeFramed}

\begin{document}


\title[Towards Directive Explanations]{Towards Directive Explanations: Crafting Explainable AI Systems for Actionable Human-AI Interactions}

\author{Aditya Bhattacharya}
\orcid{0000-0003-2740-039X}
\email{aditya.bhattacharya@kuleuven.be}
\affiliation{%
  \institution{KU Leuven}
  \city{Leuven}
  \country{Belgium}
}

\ccsdesc[500]{Human-centered computing~Human computer interaction (HCI)}
\ccsdesc[500]{Human-centered computing~Interaction design}
\ccsdesc[500]{Computing methodologies~Artificial intelligence}

\renewcommand{\shortauthors}{Bhattacharya}

\begin{abstract}
With Artificial Intelligence (AI) becoming ubiquitous in every application domain, the need for explanations is paramount to enhance transparency and trust among non-technical users. Despite the potential shown by Explainable AI (XAI) for enhancing understanding of complex AI systems, most XAI methods are designed for technical AI experts rather than non-technical consumers. Consequently, such explanations are overwhelmingly complex and seldom guide users in achieving their desired predicted outcomes. This paper presents ongoing research for crafting XAI systems tailored to guide users in achieving desired outcomes through improved human-AI interactions. This paper highlights the research objectives and methods, key takeaways and implications learned from user studies. It outlines open questions and challenges for enhanced human-AI collaboration, which the author aims to address in future work.   
\end{abstract}


\keywords{Explainable AI, Interactive Machine Learning, Explanatory Interactive Learning, Domain-Expert-AI Collaboration}


\begin{teaserfigure}
  \centering
  \includegraphics[width=0.95\linewidth]{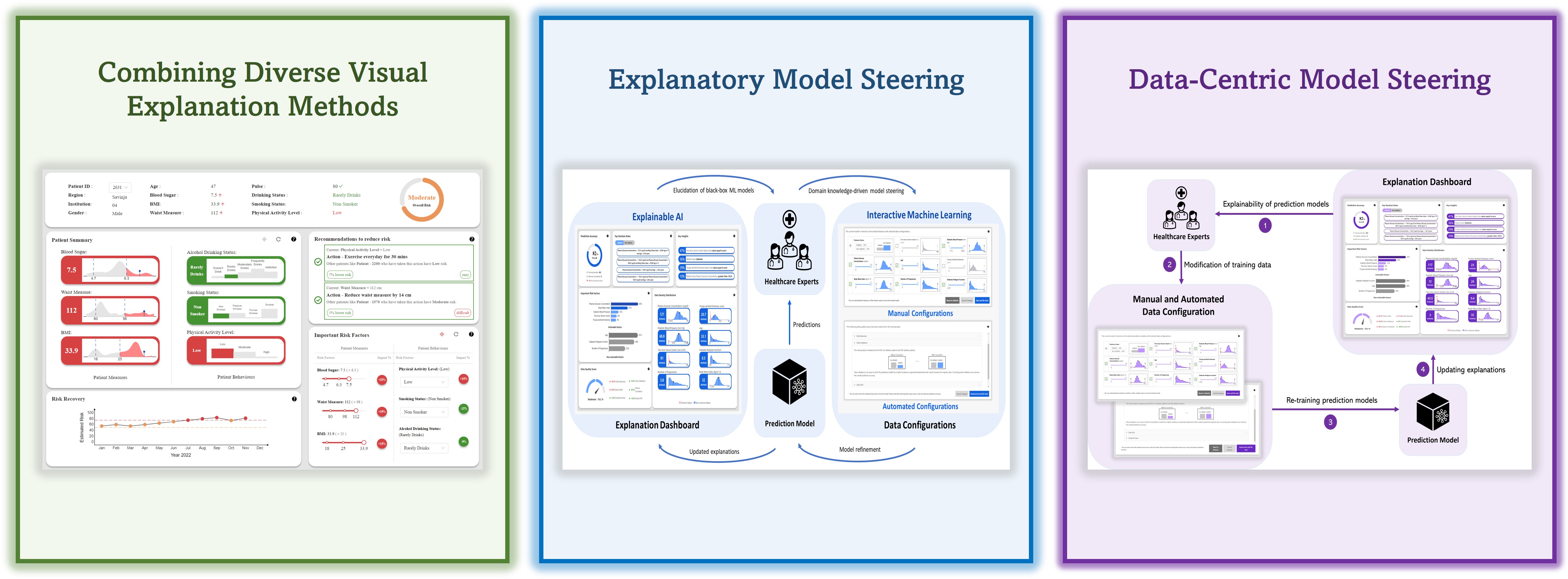}
  \caption{This research focuses on crafting Explainable AI systems for actionable human-AI interactions. Considering the current progress, the author conducted three research studies: (1) Combining diverse visual XAI methods in an explanation dashboard, (2) Explanatory model steering with domain experts through a healthcare-focused XAI system, and (3) Data-centric model steering through manual and automated configuration of training data.}
  \Description[Towards Directive Explanations]{This research focuses on crafting Explainable AI systems for actionable human-AI interactions. Considering the current progress, the author conducted three research studies: (1) Combining diverse visual XAI methods in an explanation dashboard, (2) Explanatory model steering with domain experts through a healthcare-focused XAI system, and (3) Data-centric model steering through manual and automated configuration of training data.}
  \label{fig:xil_systems}
\end{teaserfigure}

\maketitle

\begin{figure*}[]
\centering
\includegraphics[width=0.9\linewidth]{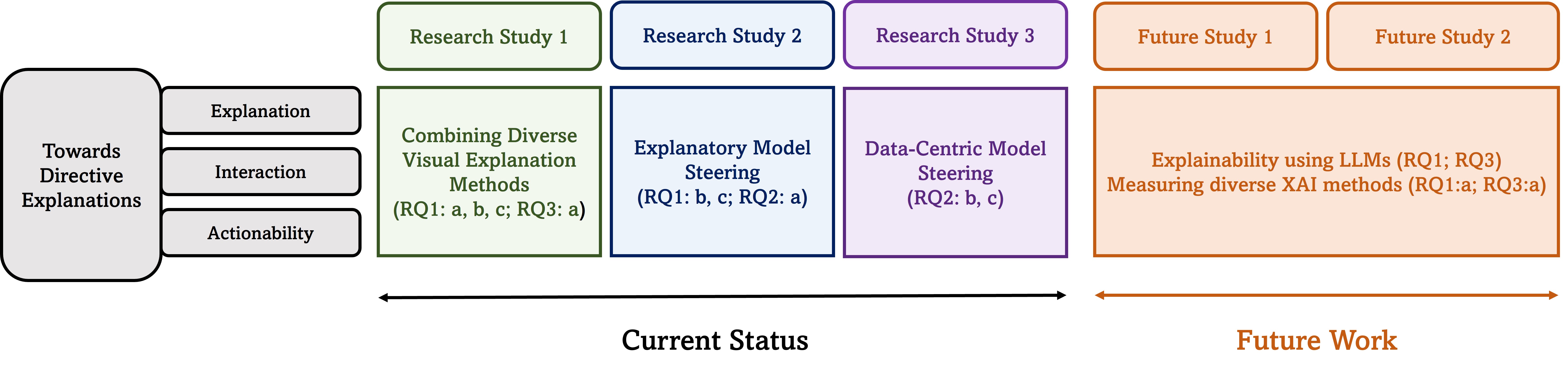}
\caption{Overview of the current research progress and planned next steps to address  research questions}
\Description[Progress overview]{Overview of the current research progress and planned next steps to address research questions}
\label{fig:progress_fw}
\end{figure*}

\section{Research Background and Motivation}
The utilisation of Artificial Intelligence (AI) systems has grown significantly in the past few years across diverse domains such as medical~\cite{pawar2020incorporating, Caruana2015, Esteva2017-pg}, finance~\cite{bove_contextualization_2022, Bussmann2021, Fahner2018}, legal~\cite{soares2019fairbydesign, wang2022pursuit, Zeng2016} and others~\cite{nti2022applications, adadi2018peeking, BhattacharyaXAI2022}. Despite the success of AI systems across various applications, the ``black-box'' nature of AI models has raised several concerns related to lack of transparency~\cite{Lim2009_why_whynot, BhattacharyaXAI2022, Miller2017} and appropriate trust~\cite{LEICHTMANN2023, stumpf2009interacting}, particularly when predicted outcomes are biased, unfair, incorrect or misguiding~\cite{brundage2018malicious, Jordan2019Artificial, Mehrabi2021}. Consequently, the field of Explainable AI (XAI) has gained much focus from AI practitioners as explanations can potentially help users to develop a clear mental model of such complex algorithms, eventually enhancing their trust for the adoption of AI systems~\cite{Guidotti2018, adadi2018peeking}. 

However, popularly adopted XAI methods~\cite{ribeiro2016why, lundberg2017unified, Islam2021XAISurvey, adadi2018peeking, BhattacharyaXAI2022} are predominantly designed for AI experts, neglecting the needs of a broader community of non-technical consumers of AI~\cite{wang_designing_2019, bove_contextualization_2022}. Moreover, prior works have questioned the efficacy of current one-off static explanations for non-expert users, i.e. users who might be experts in a particular application domain but may lack technical AI knowledge~\cite{lakkaraju2022rethinking}. Hence, there is a need for tailoring explanations for non-expert users and making XAI methods more useful, understandable, actionable and trustworthy through user-centric design principles~\cite{maxwell2023meaningful}. 

Additionally, prior works have emphasised the need for \textit{directive explanations} to non-expert users for enhanced human-AI collaborations~\cite{Singh2021directive, 2018_Spangher_Ustun}. Directive explanations can assist users in obtaining their desired predictions by interacting with the prediction system for an actionable recourse~\cite{2018_Spangher_Ustun}. Instead of \textit{static} explanations, non-expert users have expressed the need for more \textit{interactive} explanations to foster understanding and interpretation \cite{Abdul2018, Jin2020, kulesza_principles_2015} of complex AI systems. Such explanations can enable them to understand \textit{why} a certain prediction is generated and also indicate \textit{how} to obtain their desired predictions without any intervention from AI experts \cite{Lim2009_why_whynot, Singh2021_whatif, Spinner2019, Krause2016, wexler2019_whatif}. Along with interactive explanations, prior research in Interactive Machine Learning (IML) has studied the collaborative interactions of users with AI systems and their vital role in model development, fine-tuning, debugging and evaluation~\cite{fails2003, stumpf2009interacting, Amershi_Cakmak_Knox_Kulesza_2014, teso2019}. For instance, researchers of EluciDebug~\cite{kulesza_principles_2015} have found that explanation-driven interactions can facilitate users to improve prediction models by helping them to identify bugs and issues and enhance their overall mental model of the system. 

However, despite various explanation-driven interaction methods discussed in the literature~\cite{teso_leveraging_2022, teso2019, kulesza_explanatory_2010, kulesza_principles_2015, bhattacharya2023_technical_report}, researchers have identified many challenges when non-technical consumers of AI have been involved in the AI prediction process~\cite{lakkaraju2022rethinking, Schramowski2020}. For example,  non-expert users have expressed the need for more active involvement with AI models through interactive explanations of the underlying data rather than static explanations of the prediction algorithms~\cite{lakkaraju2022rethinking}. 

Moreover, recent works have emphasised the need for Data-Centric AI (DCAI)~\cite{Mazumdar2022} approaches for model improvement by improving the quality of the training data for building more reliable and trustworthy prediction models~\cite{anik_data-centric_2021}. However, prior research has shown that improving the quality of training data is not straightforward without thorough domain knowledge~\cite{feuerriegel2020fair}. Domain experts can identify certain types of biases in data to improve the training data quality. For example, instead of model developers, healthcare experts such as doctors or nurses have a better understanding of patient’s medical records, enabling them to identify biases and limitations in training data of prediction models for healthcare applications. Consequently, it is vital to capture the domain knowledge of domain experts through interactive explainability systems that give them more control over the training data~\cite{Amershi_Cakmak_Knox_Kulesza_2014, holzinger2016interactive, teso_leveraging_2022}.

\section{Research Objectives and Methods}
The author's research focuses on crafting XAI systems that provide directive explanations for non-expert users and allow them to collaborate with the system for more reliable and useful predictions. The following are the primary research objectives:
\begin{enumerate}
    \item [\textbf{O1.}] {\em Explanation:} The first objective is to compare the utility of different tailored explanation methods to elucidate predictions generated by AI models for non-expert users.
    \item [\textbf{O2}.] {\em Interaction:}  Second, the author aims to design interaction mechanisms that can be combined with different explanation methods as a basis to develop more powerful XAI systems. The effect of providing different levels of control over explanations, training data and prediction models on user understanding and trust will be investigated.
    \item [\textbf{O3}.] {\em Actionability:} Finally, the author will research approaches to guide users in obtaining their desired predictions through interactive explanations and data-centric approaches.  
\end{enumerate}

The following high-level research questions further complement the preceding research objectives:
\begin{enumerate}
\item[\textbf{RQ1}:] What is the effectiveness of different tailored explanation methods for explaining AI models to non-expert users?
\begin{itemize}
    \item[\textbf{(a)}] Are specific explanation methods better than other methods in explaining the outcomes of models? 
    \item[\textbf{(b)}] How can we combine explanation methods to develop more powerful explanation interfaces? 
    \item[\textbf{(c)}] How can explanation methods be tailored to the needs of non-expert users?
\end{itemize}
\item[\textbf{RQ2}:] How can interactive explanations facilitate non-expert users in model steering through data-centric approaches?
\begin{itemize}
    \item[\textbf{(a)}] How can different types of explanations assist non-expert users in model steering? 
    \item[\textbf{(b)}] How can non-expert users use different data-centric model steering approaches to improve prediction models? 
    \item[\textbf{(c)}] What is the impact of different levels of control for model steering?
\end{itemize}
\item[\textbf{RQ3}:] How can we tailor explanations to generate actionable insights?
\begin{itemize}
    \item[\textbf{(a)}] What is the effectiveness of different explanation methods for actionable insights? 
    \item[\textbf{(b)}] How can we design interactions to enhance the actionability of explanations?
\end{itemize}
\end{enumerate}

To address the above research questions, XAI systems are designed and developed following user-centric principles. Initially, exploratory studies are conducted to collect the needs of the target users for specific application areas. Then, in multiple iterations, low-fidelity prototypes are designed for initial evaluation with the target users. Then, high-fidelity prototypes are developed and evaluated through quantitative, qualitative or mixed-methods user studies. Finally, the data collected from these studies are analysed to extract valuable insights. These insights are instrumental in addressing the research questions and formulating design implications and guidelines for future research.

\section{Current Status}
This section describes the current progress of the author's research. To fulfil the research objectives, three research studies with two healthcare-focused interactive XAI systems have been conducted to date. The author has conducted six user studies in total involving healthcare experts and patients by following a user-centric design and development process.

\colorbox{study1}{ \textcolor{ForestGreen}{\textbf{Research Study 1}}} - In his first research study~\cite{Bhattacharya2023}, the author designed and developed an explanation dashboard for monitoring the risk of diabetes onset. This interactive dashboard provides explanations for a diabetes risk prediction model by combining data-centric, feature importance and example-based local explanations. The dashboard aims to assist healthcare experts like nurses and physicians in monitoring patients and recommending measures to minimise their risk of type-2 diabetes. First, an exploratory user study was conducted through focus group discussion followed by a co-design session with 4 registered nurses from the Community Healthcare Centre dr. Adolf Drolc in Maribor, Slovenia, to collect the user requirements and understand their challenges in monitoring diabetic patients. Next, a qualitative user study was conducted through individual interviews with 11 healthcare experts to evaluate our low-fidelity click-through prototypical dashboard. Finally, a mixed-methods user study was conducted with 45 healthcare experts and 51 diabetic patients to evaluate our high-fidelity web application prototypical explanation dashboard. Through these user studies, the author measured the usefulness, understandability, actionability and trustworthiness of different types of explanations included in the dashboard. The results underscored the importance of the author's representation of data-centric explanations that presented local explanations with a global overview of feature importance and example-based explanations. However, both healthcare experts and patients highlighted the importance of combining the different types of explanations for recommending risk-mitigating actions, indicating that any limitation of an individual explanation method can be complemented by other methods.
\colorlet{framecolor}{study1font}
\colorlet{shadecolor}{study1}
\setlength\FrameRule{0pt}
\begin{frshaded*}
The following are the key contributions of this study:
\begin{itemize}
    \item It was found that combining different XAI methods is essential to address different dimensions of explainability for a holistic explanation of predictive models. Including only one or a few types of XAI methods can only provide partial explainability. 
    \item The perspectives of healthcare experts and patients were collected to compare the different types of explanations in terms of understandability, usefulness, actionability and trustworthiness. This research collected insights about the perceived importance of directive data-centric explanations within healthcare XAI systems, illuminating a relatively unexplored subject in the XAI literature.
    \item Design implications for tailoring visually directive explanations for healthcare experts were presented considering the insights captured from this study.
\end{itemize}
\end{frshaded*}

\par\bigskip

\colorbox{study2}{ \textcolor{study2font}{\textbf{Research Study 2}}} - In the second research study~\cite{bhattacharya2024exmos, bhattacharya2023_technical_report}, the author designed and developed an Explanatory Model Steering (EXMOS) system for healthcare experts. The EXMOS system provides different types of global explanations to support healthcare experts in improving prediction models through manual and automated data configurations. To evaluate this system, two between-subject user studies were conducted: one quantitative study with 70 healthcare experts and another qualitative study with 30 healthcare experts. The impact of diffrent types of global explanations on trust, understandability and model improvement was measured in this research study. The results highlighted the inefficiency of global model-centric explanations for guiding users during the configuration of the training data. Despite the benefits of data-centric global explanations in helping users comprehend the post-configuration system changes, it was found that a hybrid fusion of both data-centric and model-centric global explanations was most effective for steering prediction models.

\colorlet{framecolor}{study2font}
\colorlet{shadecolor}{study2}
\setlength\FrameRule{1.5pt}
\begin{frshaded*}
The following are the key contributions of this study:
\begin{itemize}
    \item An XAI system that uses different types of data-centric and model-centric global explanations was designed and implemented to assist healthcare experts in sharing their domain knowledge. 
    \item Perspectives of 100 healthcare experts were collected to highlight the importance of multifaceted explanations for a holistic explainability of the system. 
    \item Guidelines for the design and implementation of explanatory model steering systems for healthcare applications were presented considering the insights collected from this study.
\end{itemize}
\end{frshaded*}

\colorbox{study3}{ \textcolor{study3font}{\textbf{Research Study 3}}} - In the latest research study, the author expanded his research with the EXMOS system to investigate the impact of different levels of control for model steering through data-centric approaches. More specifically, the latest research explored the effectiveness of manual and automated data configuration methods in healthcare-focused XAI systems. A between-subject mixed-methods user study was conducted with 74 healthcare experts to evaluate the effectiveness of different data configuration methods across multiple measures such as understandability, trust, model improvement, explanation goodness and satisfaction, feedback importance and usability and system interactions. It was found that the study participants could significantly improve the performance of the prediction model using the manual configurations instead of the automated method. The study findings highlighted the necessity of a higher involvement of domain experts for enhanced human-AI collaboration.

\colorlet{framecolor}{study3font}
\colorlet{shadecolor}{study3}
\setlength\FrameRule{1.5pt}
\begin{frshaded*}
The following are the key contributions of this study:
\begin{itemize}
    \item A user-centric design and implementation of different manual and automated data configuration approaches that enabled domain experts to share their domain knowledge and improve prediction models was presented.
    \item Detailed insights comparing the significance of manual and automated data configuration methods across multiple measures from the perspective of healthcare experts were presented. The results highlight the necessity of granting more control to domain experts for model improvement.
    \item Design guidelines for manual and automated data-centric collaborative approaches for human-AI interactions were presented considering the insights collected from the user study.
\end{itemize}\end{frshaded*}

\section{Future Work}
In his previous research studies, the author has mostly focused on different types of visual explanations designed to elucidate prediction models built on structured datasets. However, with the recent advancements in Generative AI~\cite{Sun2022GenAI} and Large Language Models (LLMs)~\cite{zhao2023explainability}, he plans to research on designing XAI systems, which include AI models built on unstructured data such as text or images. The author aims to follow a similar research procedure as our previous studies to explore different approaches for actionable and human-friendly explanations. Furthermore, his current research has highlighted the need for more objective measures to evaluate different XAI methods instead of subjective metrics. Thus, as next steps of his current research, the author wants to focus on the following areas: 

\colorbox{futurework}{ \textcolor{futureworkfont}{\textbf{Explainability using LLMs}}} - The author hypothesises that LLMs can be used to generate more natural and human-friendly explanations and foster more natural interaction approaches with AI models. More specifically, the author wants to include LLM-based chatbots~\cite{lakkaraju2022rethinking, Slack2023, sallam2023chatgpt} and allow non-expert users to interact with prediction models through conversation-based explanations and model steering approaches. He proposes to conduct randomised control experiments with non-expert users to measure the effectiveness of LLMs in generating more actionable, causal and contextual explanations. The author would also like to collect more qualitative data to understand the benefits and pitfalls of using LLMs in explanatory model steering.

\colorbox{futurework}{ \textcolor{futureworkfont}{\textbf{Measuring diverse XAI methods}}} - More recently, many researchers have questioned the efficacy of current quantitative user-centric metrics to compare different XAI methods~\cite{ALI2023_XAI_Survey, liao2022humancentered, Nauta2023}. Most of these metrics are highly subjective to the research participants and do not provide a generalised evaluation of different types of XAI methods for other users or applications. Thus, the author's current research studies have raised a vital open question: ``\textit{how to measure diverse XAI methods in a unified and fair approach as each method elucidates different aspects of an AI system?}" Therefore, he proposes to conduct a systematic review of the research literature to compare the benefits and drawbacks of subjective user-centric XAI metrics and objective XAI metrics. After this systematic review, he aims to formulate a unified and fair evaluation approach for inter-system and intra-system comparisons of XAI methods from this research and evaluate this novel approach with XAI researchers to collect their feedback.

\begin{acks}
The author would like to thank his supervisor, Prof. Katrien Verbert, for her constant support throughout his PhD. This work is supported by research grants: Research Foundation–Flanders (FWO, grant G0A3319N, G0A4923N, G067721N) and KU Leuven Internal Funds (grant C14/21/072). He would like to extend his gratitude to his collaborators Dr. Simone Stumpf (University of Glasgow, UK), Dr. Gregor Stiglic (University of Maribor, Slovenia) and the Augment HCI group, KU Leuven, for their support during this research work.
\end{acks}

\bibliographystyle{ACM-Reference-Format}
\bibliography{references}




\end{document}